\documentclass{aa}
\usepackage[varg]{txfonts}
\usepackage{graphicx}
\usepackage{txfonts}
\usepackage{natbib}
\usepackage{verbatim}
\usepackage[pdftex]{color}
\bibpunct{(}{)}{;}{a}{}{,} % to follow the A&A style

\usepackage{booktabs}
\definecolor{ultramarine}{rgb}{0.07, 0.1, 0.6} 
\definecolor{myblue}{rgb}{0.07, 0.2, 0.6} 
\definecolor{dopal}{rgb}{.70, .25, .05}

\begin{document}
\title{The pulsational properties of ultra-massive DB white dwarfs with carbon-oxygen cores coming from single-star evolution}

\author{Alejandro H. C\'orsico \inst{1,2}, 
 Leandro G. Althaus \inst{1,2},
 Pilar Gil Pons\inst{3}, and  
 Santiago Torres\inst{3,4}}
\institute{Grupo de Evoluci\'on Estelar y Pulsaciones. 
           Facultad de Ciencias Astron\'omicas y Geof\'{\i}sicas, 
           Universidad Nacional de La Plata, 
           Paseo del Bosque s/n, 1900 
           La Plata, 
           Argentina
           \and
           CCT - CONICET
           \and
           Departament de F\'\i sica, 
           Universitat Polit\`ecnica de Catalunya, 
           c/Esteve Terrades 5, 
           08860 Castelldefels, 
           Spain
           \and
           Institute for Space Studies of Catalonia, 
           c/Gran Capita 2--4, 
           Edif. Nexus 104, 
           08034 Barcelona, 
           Spain
           }
\date{Received ; accepted }

% context
\abstract{Ultra-massive white dwarfs are 
relevant for many reasons: their role as type Ia Supernova progenitors, 
the occurrence  of physical processes in the  asymptotic giant-branch phase, 
the  existence of high-field magnetic white dwarfs,  and the occurrence of double 
white dwarf mergers. Some hydrogen-rich ultra-massive white dwarfs are pulsating stars, 
and as such, they offer the possibility  of studying their interiors through 
asteroseismology. On the other hand,  pulsating helium-rich ultra-massive white dwarfs 
could be  even more attractive objects for asteroseismology if they were found,
as they should be hotter and less crystallized than pulsating hydrogen-rich white dwarfs, 
something that would pave the way for probing their deep interiors.} 
% Aims
{We explore the pulsational properties of ultra-massive helium-rich white dwarfs with carbon-oxygen and oxygen-neon cores resulting from single stellar evolution. 
Our goal is to provide a theoretical basis that could eventually help
to discern the core composition of ultra-massive white dwarfs and the 
scenario of their formation through asteroseismology, anticipating the possible 
future detection of pulsations in this type of stars.}
% Methods
{We focus on three scenarios for the formation of ultra-massive white dwarfs. 
First, we consider stellar models coming  from two recently proposed 
single-star evolution scenarios for the formation of ultra-massive 
white dwarfs with carbon-oxygen cores that involve rotation of  
the degenerate core after core helium burning, and  reduced  
mass-loss rates in massive asymptotic giant-branch stars. Finally, we 
contemplate ultra-massive oxygen-neon core white-dwarf models resulting from
standard single-star evolution. 
We compute the adiabatic pulsation gravity-mode periods for models in a range of effective temperatures embracing the instability strip of average-mass pulsating 
helium-rich white dwarfs, and compare the characteristics of 
the mode-trapping properties for models of different formation scenarios 
through the analysis of the period spacing.}
% Results
{We find that, given that the white dwarf models coming from the three 
scenarios considered are characterized by distinct core chemical profiles, their pulsation 
properties are also different, thus leading to distinctive signatures in the 
period-spacing and mode-trapping properties.}
% Conclusions
{Our results indicate that, in case of an eventual detection of pulsating 
ultra-massive helium-rich  white dwarfs, it would be possible to derive valuable information encrypted in the core of these stars in connection with the origin of 
such exotic objects. This is of utmost importance regarding recent evidence
for the existence of a population of ultra-massive WDs with carbon-oxygen cores. 
The detection of pulsations in these stars has many chances to be achieved 
soon through observations collected with ongoing space missions.}

\keywords{stars:  evolution  ---  stars: interiors  ---  stars:  white
  dwarfs --- stars: pulsations}
\titlerunning{Pulsational properties of ultra-massive DB white dwarfs}
\authorrunning{C\'orsico et al.}

\maketitle

\section{Introduction}
\label{introduction}

Massive  white-dwarf  (WD) stars have become  a topic of great interest
in recent years, as their origin and evolutionary properties are key
to  understand  type  Ia
Supernova,  the occurrence  of  physical processes  in the  Asymptotic
Giant  Branch  (AGB) phase,  the  theory  of crystallization  and  the
existence  of  high-field  magnetic  WDs,  as  well  as  to  help  our
understanding          of          double          WD          mergers
\citep{2015ASPC..493..547D,2020A&A...638A..93R}. 
The mass distribution of the entire WD population is clearly peaked at 
$M_{\star} \sim 0.6 M_\sun$, but massive WDs also show a well-defined peak at 
$M_{\star} \sim 0.82 M_\sun$
\citep[e.g.][]{2013ApJS..204....5K,Jimenez2018}, which has been attributed, among other factors, 
to the empirical crystallisation delay at higher masses observed in {\it Gaia} 
\citep{2019MNRAS.482.5222T,2019Natur.565..202T,2020ApJ...898...84K}.
In  addition,  the   existence   of
ultra-massive  WDs  ($M_{\star}  \gtrsim 1  M_\sun  $)  has  been
reported               in               numerous               studies
\citep{2010MNRAS.405.2561C,2013MNRAS.430...50C,2013ApJ...771L...2H,
  2016MNRAS.455.3413K,2017MNRAS.468..239C,2018ApJ...861L..13G,
  2020NatAs...4..663H,2020ApJ...901...93B}.

The  standard scenario for the formation of ultra-massive WD involves 
 single progenitor stars of initial masses higher than 6--9 $M_\sun$ 
(the precise value depends strongly on the  input physics and metallicity) that experience off center 
carbon (C)  burning  during the  Super  AGB when the carbon-oxygen (CO) core mass has
  grown to about 1.05$\,M_\sun$, thus  eventually leading to the formation of 
ultra-massive WDs with oxygen-neon (ONe)       cores       
(UMONe WDs; see,      e.g. \citealt{garciaberro1994};
\citealt{2007A&A...476..893S, 2017PASA...34...56D}). An alternative scenario
involves the double WD merger. 
Theoretical computations indicate that double WD mergers would contribute 
to some an extent to the single  massive WD population \citep{2017A&A...602A..16T,2018MNRAS.476.2584M}.
In particular, \cite{2020A&A...636A..31T} predict that a fraction of  
all observable single WDs more massive than 0.9$\,M_\sun$ within 100  pc  might  result    
mostly    from    the   merger    of    two    WDs. Also, \cite{2020ApJ...891..160C} conclude
that  about   20  $\%$  of  massive and  ultra-massive 
WDs  in  the   mass  range
0.8--1.3$\,M_\sun$  results  from  double-WD  mergers. 

Evidence from GAIA kinematics of large cooling delays of ultra-massive WDs on the Q-branch reported by \cite{2020ApJ...891..160C} and the location of
the Q-branch on the color-magnitude diagram  are consistent with a population
of ultra-massive WDs with CO cores (UMCO)  \citep[see also][]{2019Natur.565..202T,2020ApJ...902...93B}.
 Given  recent studies based on post merger evolutionary calculations that suggest 
 the formation of an ONe core after WD merger in ultra-massive WDs \citep{2012ApJ...748...35S,2020arXiv201103546S},
the existence of a population of UMCO WDs is difficult to understand. 
Recently, \cite{2020arXiv201110439A} have  explored single-evolution  scenarios that
could  lead to  the formation of  UMCO WDs. These authors
have  studied  the  evolutionary  and pulsational  properties  of  the
resulting DA WDs and compared them with those of ultra-massive WDs with ONe cores,
thus establishing  a theoretical basis  that could eventually  help to
infer the  core composition of  ultra-massive WDs and the  scenario of
their  formation. Specifically,  \cite{2020arXiv201110439A} 
study  two possible  single
evolution  scenarios for  the  formation of  UMCO  WDs.  One  scenario
exploits   wind rates  and convective boundary  uncertainties  during   the  thermally-pulsing
AGB (TP-AGB)  phase,
and involves  the reduction of these  rates below the values  given by
standard  prescriptions \citep[see, e.g.,][]{decin2019}. The  other scenario requires
the occurrence  of rotation in  degenerate CO cores, naturally expected as a consequence of core contraction at the end  of  core  helium (He) exhaustion
\citep{dominguez1996}.  \cite{2020arXiv201110439A} show  that both  the evolutionary  and pulsational properties of the UMCO WDs formed through these two single
evolution  scenarios are  markedly different  from those of 
UMONe  WDs. 
Such differences in evolutionary and pulsational properties may eventually be used to shed light on
%discern 
the core composition of ultra-massive WDs.

The  chemical stratification and internal  structure of pulsating WDs, and in particular,  pulsating ultra-massive WDs,  can be in principle probed  by means of asteroseismology  
\citep{2008ARA&A..46..157W,2008PASP..120.1043F,2010A&ARv..18..471A,2019A&ARv..27....7C}. Indeed, several ultra-massive H-rich  WDs  (DA  WDs)  exhibit  $g$(gravity)-mode  
pulsational instabilities \citep{2005A&A...432..219K,2010MNRAS.405.2561C,
2013MNRAS.430...50C, 2013ApJ...771L...2H,2017MNRAS.468..239C,
2019MNRAS.486.4574R}, and they are part of the ZZ Ceti (or DAV) class  
of variable H-rich 
WDs. A recent attempt to explore the internal 
structure of ultra-massive ZZ Cetis stars via asteroseismology  
has been conducted by \cite{2019A&A...632A.119C}.
These authors emphasize the need for the detection of more periods 
and more pulsating ultra-massive WDs, something that could soon be 
achieved with observations from space, such as those of the Transiting 
Exoplanet Survey Satellite \citep[{\it TESS};][]{2015JATIS...1a4003R}. 

To explore the internal structure of pulsating WDs 
through asteroseismology, it is crucial to employ stellar models with 
detailed chemical profiles resulting from the complete evolution of their 
progenitor stars, as well time-dependent element diffusion during the 
WD evolution. Indeed, the details of the shape of the internal chemical 
profiles of WDs are key in relation to the characteristics of the $g$-mode pulsation spectrum of these stars, and in particular, to the mode trapping properties. 
This was realized almost two decades ago by the La Plata Group\footnote{\tt http://fcaglp.unlp.edu.ar/evolgroup} 
in the case  of DAV stars \citep[e.g.,][]{2001A&A...380L..17C}, 
pulsating He-rich atmosphere WD  stars, 
called DBV or V777 Her stars \citep[e.g.,][]{2004A&A...417.1115A}, 
and  pulsating PG 1159 stars, also called GW Vir  stars \citep[e.g.,][]{2006A&A...454..863C}.

In this paper we extend the scope of the study of \cite{2020arXiv201110439A} 
by exploring the pulsational  properties of the ultra-massive  WDs with 
He-rich
atmospheres (DB WDs)  resulting from  single-star evolution.
In the case of H-deficient WDs, the mass  distribution 
shows an apparent deficiency  of ultra-massive objects
\citep{2019MNRAS.482.5222T,2019MNRAS.486.2169K}. 
Despite that ultra-massive H-deficient WDs are not so common 
as ultra-massive H-rich WDs, several observational evidences are found. Indeed, 
some recent works point to the existence of a small sub-population 
of ultra-massive DO (He-rich atmospheres with ionization lines) 
and DB WDs \citep{2014A&A...572A.117R,2020ApJ...901...93B}. Another
class of
H-deficient ultra-massive WDs are the DQ WDs, 
that show He and C at their atmospheres \citep{2013ApJS..204....5K,2020A&A...635A.103K}.
In particular, \cite{2014A&A...572A.117R} found one ultra-massive DO WD  
from the data release 10 (DR10) of the Sloan Digital Sky Survey 
\citep[SDSS;][]{2000AJ....120.1579Y} 
with a mass of $1.07 M_{\odot}$ if it is an ONe-core WD,
or $1.0 9M_{\odot}$ if it is a CO-core WD. On the other hand,
\cite{2020ApJ...901...93B} find five ultra-massive DO 
WDs stars\footnote{Actually, two DO, two DOZ (atmospheres with traces of 
metals) and one DOA (atmospheres with H lines) 
WDs \citep{2020ApJ...901...93B}.} from DR12  of the SDSS, with masses 
in the range $1.01 \leq M_{\star}/M_{\odot} \leq 1.06$. 
\cite{2019ApJ...880...75R}   have  reported   the   existence  of   an
ultra-massive  DB  WD   with  no  H  lines  in   a  young  open
cluster. Its  effective temperature, in  excess of $25,000$ K, places
this ultra-massive DB  WD inside the DBV instability  strip. 
Recently, \cite{2020MNRAS.499L..21P} discovered an ultra-massive WD
with He-rich atmosphere and traces of H (DBA spectral class) with 
$T_{\rm eff}= 31\,200\pm1200$ K and $M_{\star}= 1.33 M_{\odot}$
that exhibits photometric variability with a single period of $353.456$ s
that could be due to fast rotation, supporting a merger scenario 
for its formation. However, it cannot be completely 
ruled out that the variability is due to pulsations, since
it is close to the blue edge of the DBV instability strip.
We note that a significant number of ultra-massive He-atmosphere WDs are magnetic, including the one in \cite{2019ApJ...880...75R} and 
also the hot DQ WDs \citep[see, e.g.,][]{2013ASPC..469..167D}. The presence of a strong magnetic field could complicate the study of pulsations, since an intense magnetic field  would be capable of inhibiting convection and therefore have a dramatic 
effect on the driving mechanism of the pulsations \citep{2015ApJ...812...19T}. 

In this paper, we will show that pulsational properties of  
the ultra-massive DB WDs are much
more strongly dependent  on their formation scenario than  in the case
of the  DA WD ones studied in \cite{2020arXiv201110439A}. This is 
due to the fact that pulsating DB WDs  
are much hotter than ZZ Cetis, and therefore, their cores are 
less crystallized.  As a result, $g$-mode pulsations can penetrate much 
deeper in the star, thus carrying valuable information about the 
core chemical structure and composition. Hence, the eventual  detection of
pulsating ultra-massive DB WDs will constitute a much clear and unique
opportunity to discern the core composition and origin of 
ultra-massive WD population in general.

The paper is  organized as follows.  In  Sect.~\ref{codes} we describe
the  stellar  codes employed to   compute  the  evolutionary  and  pulsational
properties  of our  DB WD  models,  and briefly  summarize the  single
stellar evolution scenarios that lead to our initial
UMCO WD models.  In Sect.~\ref{pulsation_results} we describe
the  pulsational  properties  of   the  resulting  WDs.   Finally,  in
Sect.~\ref{conclusions} we summarize the main findings of the paper.

\section{Numerical codes and ultra-massive WD models}
\label{codes}

The evolutionary and  pulsational properties of our UMCO and UMONe DB WD 
models
were computed  with the same stellar codes we  used in \cite{2020arXiv201110439A}. For  the pulsational  analysis we used  the {\tt  LP-PUL} pulsation
code described in \citet{2006A&A...454..863C}. Relevant for this work,
the code  adopts the "hard-sphere" boundary conditions to  account for
the  effects   of  crystallization   on  the  pulsation   spectrum  of
$g$ modes.  These   conditions  assume  that  the   amplitude  of  the
eigenfunctions  of  $g$-modes   is  null  below the  solid/liquid
interface because of  the non-shear modulus of the  solid, as compared
with  the fluid  region \citep[see][]{1999ApJ...526..976M}.  The inner
boundary   condition for pulsations  corresponds   to    
the   mesh-point   at   the crystallization front
\citep[see][]{2004A&A...427..923C,2005A&A...429..277C,2019A&A...621A.100D,2019A&A...632A.119C}. The    Brunt-V\"ais\"al\"a    frequency     is    computed    as    in
\cite{1990ApJS...72..335T}.  The  computation of  the Ledoux  term $B$
includes the effects of having  multiple chemical species that vary in
abundance. For the  computation of the WD evolutionary  models  we
used the  {\tt LPCODE} stellar  evolutionary code that has  been widely
used and tested in numerous  stellar evolutionary contexts of low-mass
and         WD          stars         \citep[see][for         details]
{2003A&A...404..593A,2005A&A...435..631A, 2013A&A...555A..96S, 2015A&A...576A...9A,
  2016A&A...588A..25M,2020A&A...635A.164S,
  2020A&A...635A.165C}. In particular, the treatment
of crystallization is  based on  phase diagrams of  \cite{2010PhRvL.104w1101H} for
dense  CO mixtures,  and  that of  \cite{2010PhRvE..81c6107M} for  ONe
mixtures.    {\tt  LPCODE}   considers  a  new
full-implicit  treatment  of  time-dependent  element  diffusion  that
includes  thermal and  chemical diffusion  and gravitational  settling
\citep{2020A&A...633A..20A}. In contrast to \cite{2020arXiv201110439A},  
we have considered in this work the effect of Coulomb  separation of
ions \citep{2010ApJ...723..719C,2013PhRvL.111p1101B,2020A&A...644A..55A}.
As a result of this Coulomb diffusion,  ions with  larger $Z$   
move to  deeper layers.  Coulomb diffusion is not negligible in the 
dense C- and He-rich envelopes of  ultra-massive WDs,  thus  preventing  the inward
diffusion of  He toward  the core  and leading  to a  much sharper
transition region.

We summarize now  the main formation scenarios of our  UMCO WD models.
Full details can be found in  \cite{2020arXiv201110439A}.
These authors explored two different single-evolution scenarios 
that can lead to the formation of UMCO WDs. One of  them involves a 
reduction in the  mass-loss rates usually
adopted for the evolution of massive AGB stars. They showed that in this
case, if the minimun CO-core mass for the occurrence of C burning
is not reached before the TP-AGB phase,  a UMCO WD of mass larger than
$M_{\star}  \gtrsim 1.05\,M_\sun $ can  be formed as a  result of the
slow growth  of the CO-core  mass during  the TP-AGB  phase.  
For  this to occur, the  mass-loss rates of  massive AGB stars  need to be  at 
least $5-20$ times lower than standard mass-loss rates when a search algorithm for 
convective neutrality is used to determine convective boundaries. 
Such treatment favored very efficient third dredge-up, which hampered core growth during the TP-AGB. On the other hand, when the strict  Schwarzschild criterion is used for the determination of convective boundaries, core growth is faster, and the required reduction of standard mass-loss rates drops to a factor of 2. The
reduction in mass-loss rates necessary to form UMCO WDs cannot be discarded and is in line with
different  pieces  of  recent  observational  evidence  that  indicate
mass-loss   rates   lower   than    expected   from   current   models
\citep[e.g.,][]{decin2019}.

The other single evolution scenario that leads to the formation of UMCO
WDs involves rotation  of the degenerate core that  results after core
He burning at the onset of the  AGB phase \citep{dominguez1996}.
As these authors, \cite{2020arXiv201110439A} find  that  the   lifting  effect  
of  rotation delays the occurrence of the second dredge-up and maintains the maximum 
temperature at lower values than that required for off-center  C ignition.    Hence,  C burning is prevented and the mass  of the resulting CO
core is larger than that at which C  burning is expected in the 
absence  of rotation. As a
result, the mass  of the degenerate CO core will  be larger than $1.05
M_\sun$ before the TP-AGB sets in. \cite{2020arXiv201110439A}  find that UMCO WDs can  be formed  even for  very  low rotation  rates, and that the range  of
initial  masses  leading to  UMCO  WDs  widens  as the  rotation  rate
increases,  whereas  the  range  for the  formation  of  ONe-core  WDs
decreases significantly.

Finally, we have considered UMONe WDs resulting
from  off-center   C-burning  during  the  single   evolution  of the
progenitor  star.  Such WD  models  were  studied in  detail  by
\cite{2019A&A...625A..87C}.   The   progenitors  of  our  UMONe WDs
experienced a violent C-ignition phase followed by the development
of an inward-propagating convective flame  that transforms the CO core
into a  degenerate ONe one.  In this scenario, ultra-massive  WDs with
stellar  masses larger  than $M_{\star}  \gtrsim 1.05  M_\sun $  and
composed  of $^{16}$O  and  $^{20}$Ne ---with  traces  of $^{23}$Na  and
$^{24}$Mg---     are          expected           to          emerge
\citep{2007A&A...476..893S,2010A&A...512A..10S}.

\begin{figure}
        \centering
        \includegraphics[width=1.\columnwidth]{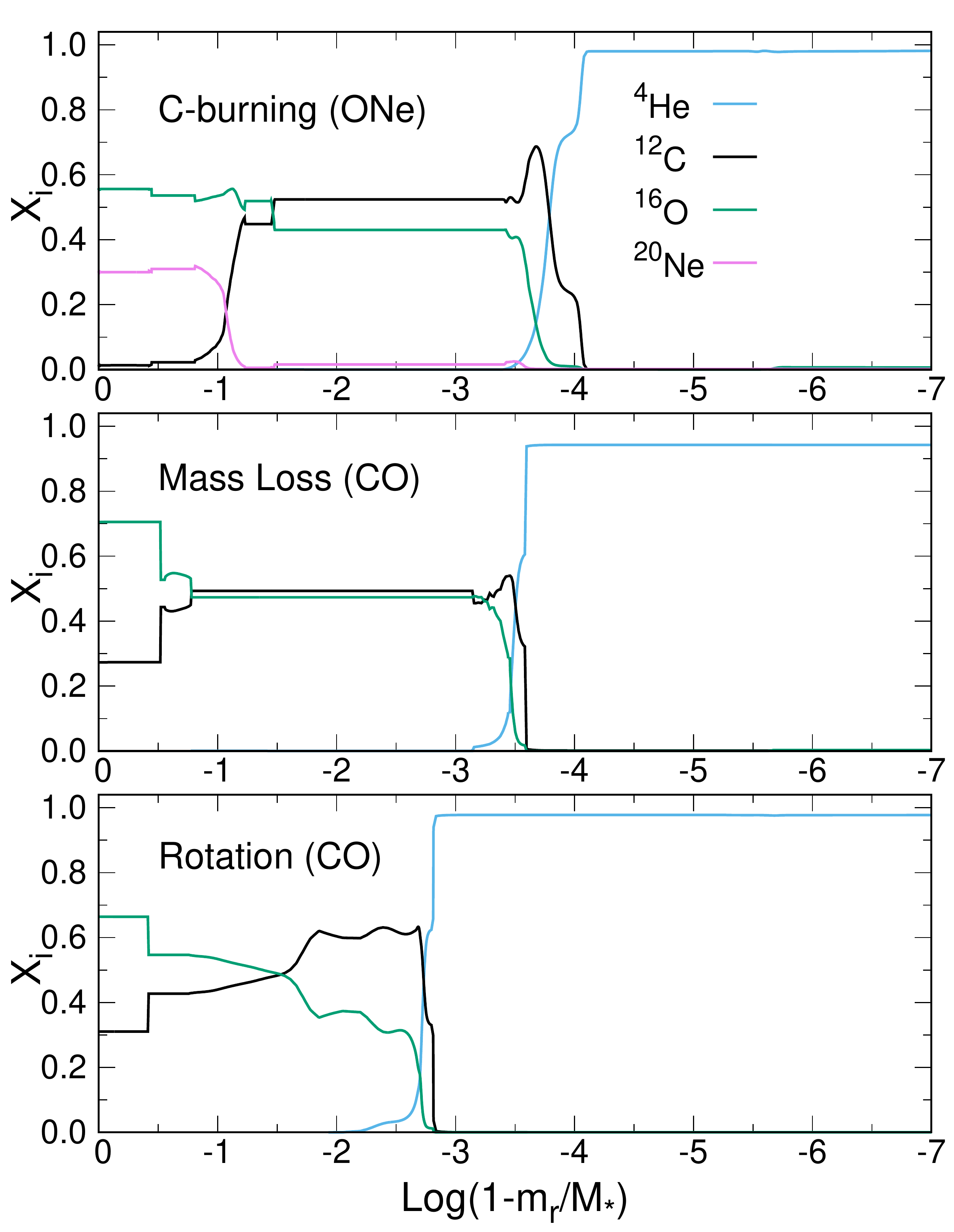}
        \caption{Abundance by mass of $^4$He, $^{12}$C, $^{16}$O, 
        and $^{20}$Ne  versus the outer mass coordinate  for the  
        $1.156 M_{\odot}$ DB WD models resulting from the evolutionary 
        scenarios we studied. From top to bottom, we show the ONe WD model from \cite{2019A&A...625A..87C},  
        the CO WD model resulting from reduced 
        mass loss during progenitor evolution, and  the CO WD model implied by 
        rotation. } 
        \label{perfiles-inicial.eps}
\end{figure}

Our starting  ultra-massive DB  WD configurations were  extracted from
the ultra-massive DA WDs of  stellar mass 1.159\,$M_{\sun}$ studied in
\cite{2020arXiv201110439A}, and formed through  the three scenarios previously 
mentioned, to which  we  have simply  removed  the  whole  H content  at  the
beginning  of   the  WD cooling   track.   Evolution  of   the  resulting
ultra-massive DB WD  configurations was followed down to  the red edge
of the DBV  instability strip, i.e., at $T_{\rm eff}\sim  20\,000$ K, in a
self-consistent way with the changes in the  internal chemical distribution
that result from the mixing of  all the core-chemical components induced by
the  mean molecular  weight  inversion left by progenitor evolution, element  diffusion, and  phase
separation of  core-chemical  constituents upon  crystallization.  The
chemical profiles of the  resulting ultra-massive DB WD configurations
are  shown   in  Fig.   \ref{perfiles-inicial.eps}.   The   top  panel
illustrates the chemical  profile for the ONe-core WD,  the second and
third  panels depict,  respectively, the  chemical profiles  resulting
from  reducing the  mass-loss rates  of an  initially $7.8  M_{\odot}$
progenitor and from  considering core rotation in the AGB  phase of an
initially  $7.6 M_{\odot}$  progenitor.   
The chemical  profiles correspond to ultra-massive  DB WD models
at the  onset of  their cooling  phase prior to  the onset  of element
diffusion and  after the core mixing  induced by the inversion  of the
mean molecular weight.  The chemical
structure  of  both  the  core  and  the  envelope  of  the  resulting
ultra-massive WDs  strongly depends on the  evolutionary scenario that
leads to their formation. In particular, because the lower  temperature and pressure 
prompted  by core rotation
favouring the formation of a less degenerate core, the He content of
the UMCO  WD resulting from  rotation is larger  than for the  UMCO WD
resulting from  reduced mass-loss rates  and  the UMONe  WD \citep{2020arXiv201110439A}.

\begin{figure}
        \centering
        \includegraphics[width=1.0\columnwidth]{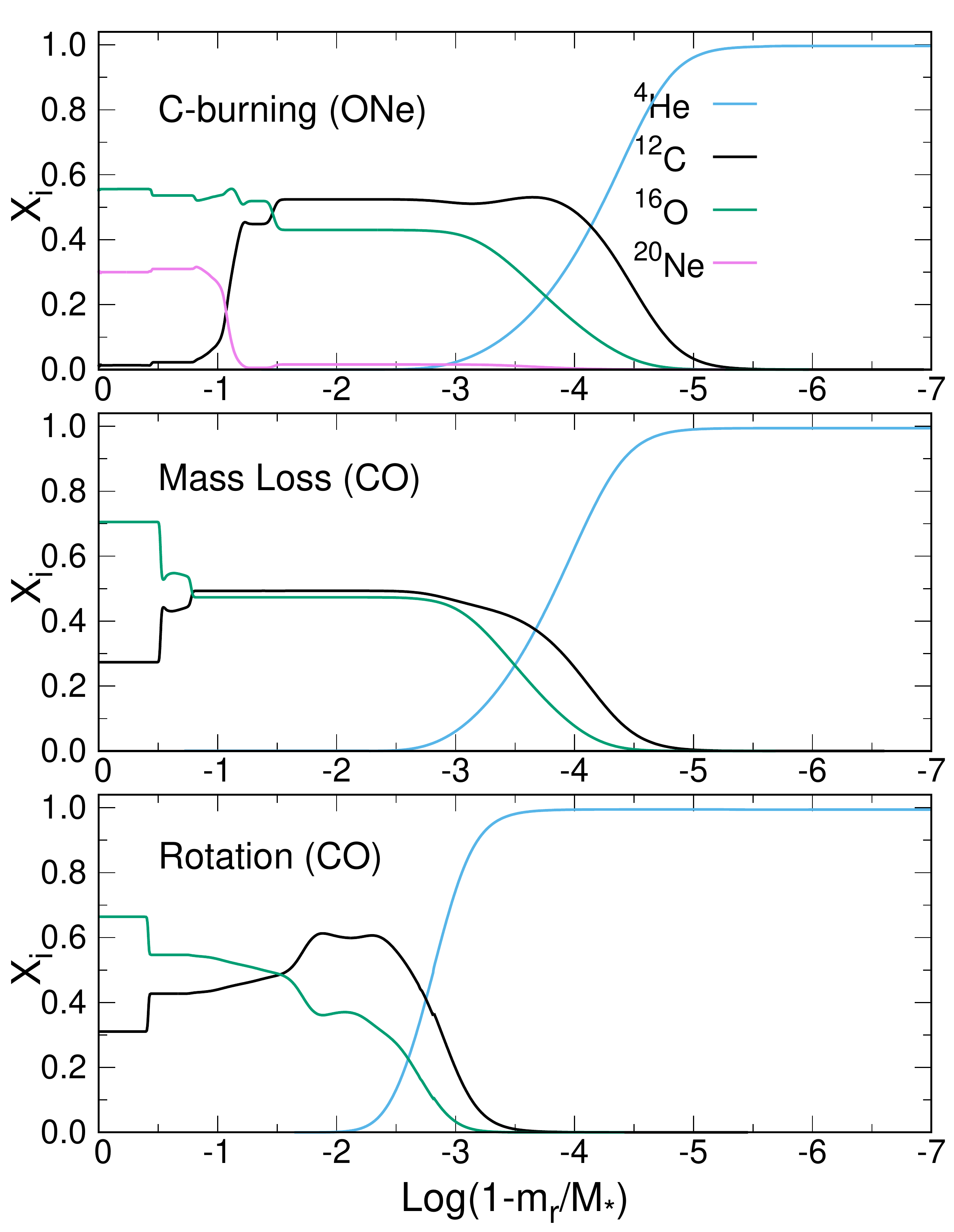}
        \caption{Same as Fig. \ref{perfiles-inicial.eps}, but at an effective 
        temperature of $T_{\rm eff}\sim  30\,000$ K, close to the 
        blue edge of the instability domain of DBV WDs. The stellar mass of the 
        models is $M_{\star}= 1.156 M_{\odot}$.} 
        \label{perfiles_30000.eps}
\end{figure}

\begin{figure}
        \centering
        \includegraphics[width=1.0\columnwidth]{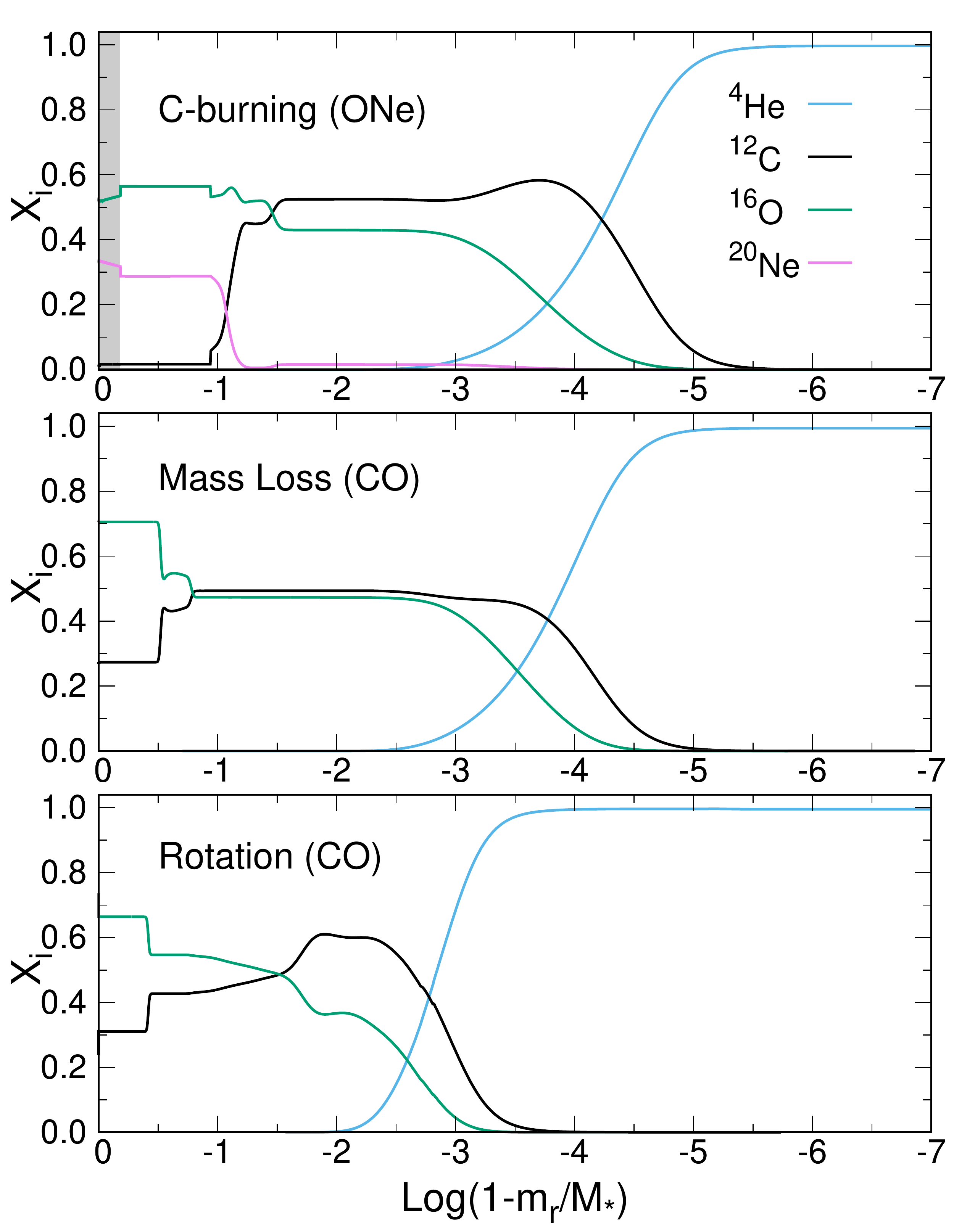}
        \caption{Same as Fig. \ref{perfiles_30000.eps}, but at an effective temperature
          of $T_{\rm eff}\sim  20\,000$ K, close to the red edge of the instability 
          domain of DBV  WDs.  The gray area in the upper panel indicates the domain of core crystallization. The percentage of crystallized mass is $\sim 34\%$.} 
        \label{perfiles_20000.eps}
\end{figure}

During WD  cooling, the  internal chemical  distribution of  our models
changes  mainly  due to  element  diffusion  and  as  a  result  of  phase
separation of the core chemical  constituents upon crystallization. By the
time evolution has proceeded to the blue (hot)  edge of DBV instability strip,
i.e., at $T_{\rm eff} \sim  30\,000$ K, chemical diffusion has strongly
smoothed out the rather abrupt initial He/C transition of our WD models 
(see Fig. \ref{perfiles_30000.eps}).  
Fig. \ref{perfiles_20000.eps}  shows the chemical profiles  at the red
(cool) edge of  DBV instability strip,  i.e., at $T_{\rm  eff} \sim 
20\,000$  K. As evolution  proceeds along  the  instability  strip, 
element  diffusion
barely modifies the  chemical profiles. For  the stellar mass
value  we  considered,   only  the UMONe  WD   sequence  develops  core
crystallization while the WD model evolves along the instability 
strip.  This  is  due  to the  larger  Coulomb  interactions
prevailing   at   the    ONe   core,   as   compared    with   the   CO
core. Crystallization for this sequence starts at $T_{\rm eff} \sim 24,500$ K, 
and by the
time  it abandons  the  instability  strip,  the mass  of  the
crystallized core amounts to about $34\%$.  The shape of the
chemical  profile  is modified  by  crystallization  not only  in  the
crystallized ONe  core left behind,  but also in liquid  regions beyond
the crystallization  front. These changes in the core composition at the liquid regions are expected to  impact the  theoretical pulsational spectrum of ultra-massive WDs 
\citep[see][]{2019A&A...621A.100D}. 

In order to explore the dependence of the pulsational properties of our models of 
ultra-massive DB WDs with the stellar mass, 
we generated an additional sequence of $M_{\star}= 1.29 M_{\odot}$ for 
each scenario by artificially scaling the mass value of each 
$1.156 M_{\sun}$ sequence
at high luminosities. In this case, due to the higher mass, all the model sequences 
experienced core crystallization as they were evolving along the instability domain of 
DBVs. In particular, at the blue edge of the instability strip, the mass of the crystallized 
core amounts to about $54\%$ in the UMONe WD model, $26\%$ for the 
UMCO WD model resulting from reduced mass loss, and $22\%$ in the 
case of the UMCO WD model implied by rotation.
These percentages change to $93\%$, $79\%$, and $79\%$, respectively,  
by the time the models reach the red boundary of the instability strip.

\begin{figure}
        \centering
        \includegraphics[width=1.0\columnwidth]{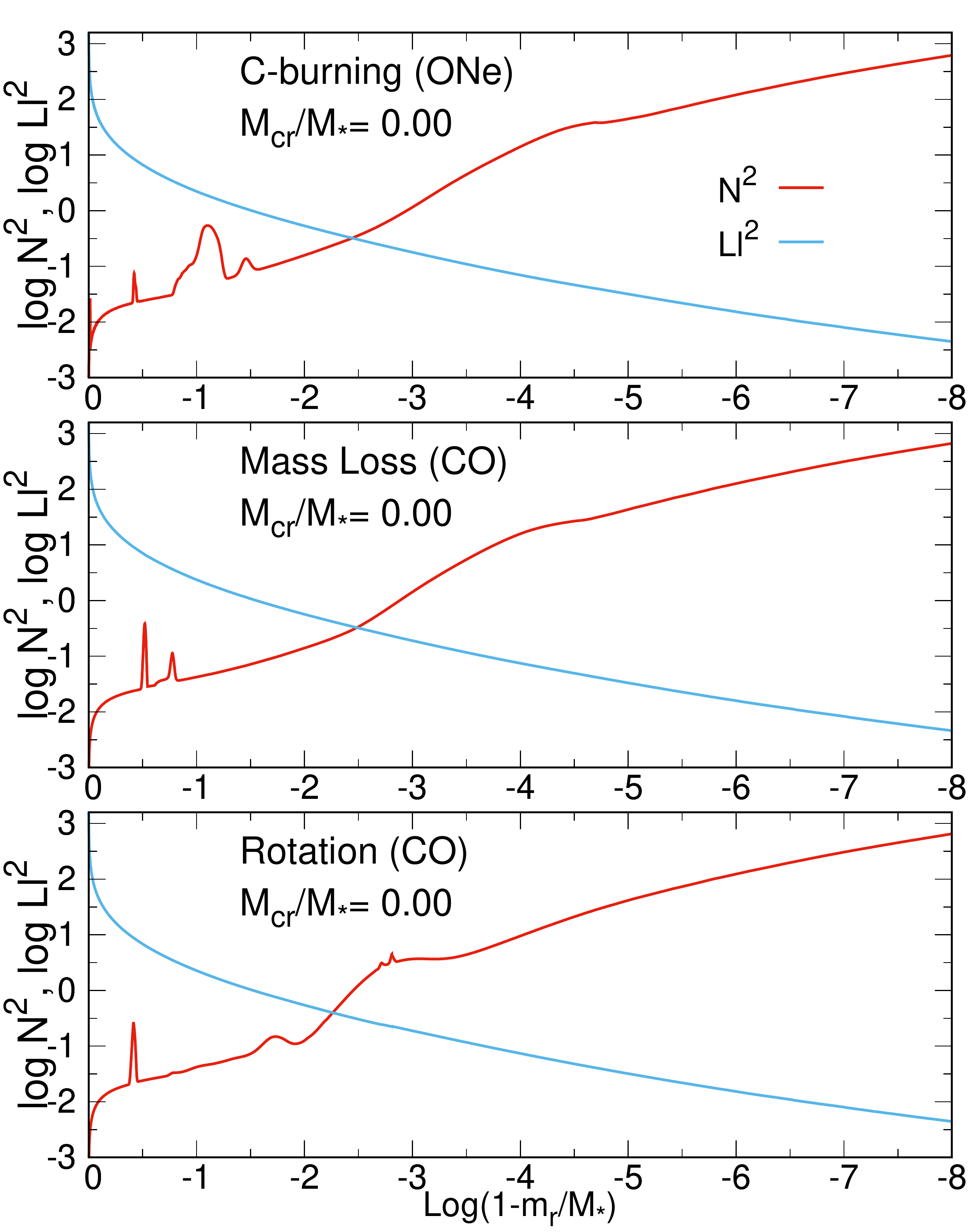}
        \caption{Logarithm of the squared Brunt-V\"ais\"al\"a and Lamb 
frequencies (red and blue lines respectively) corresponding to the 
same models analyzed  in Fig. \ref{perfiles_30000.eps}, characterized by
$M_{\star}= 1.156 M_{\odot}$  and $T_{\rm eff}\sim 30\,000$ K. 
The Lamb frequency corresponds to dipole ($\ell= 1$) modes.} 
        \label{bvf_30000.eps}
\end{figure}

\begin{figure}
        \centering
        \includegraphics[width=1.0\columnwidth]{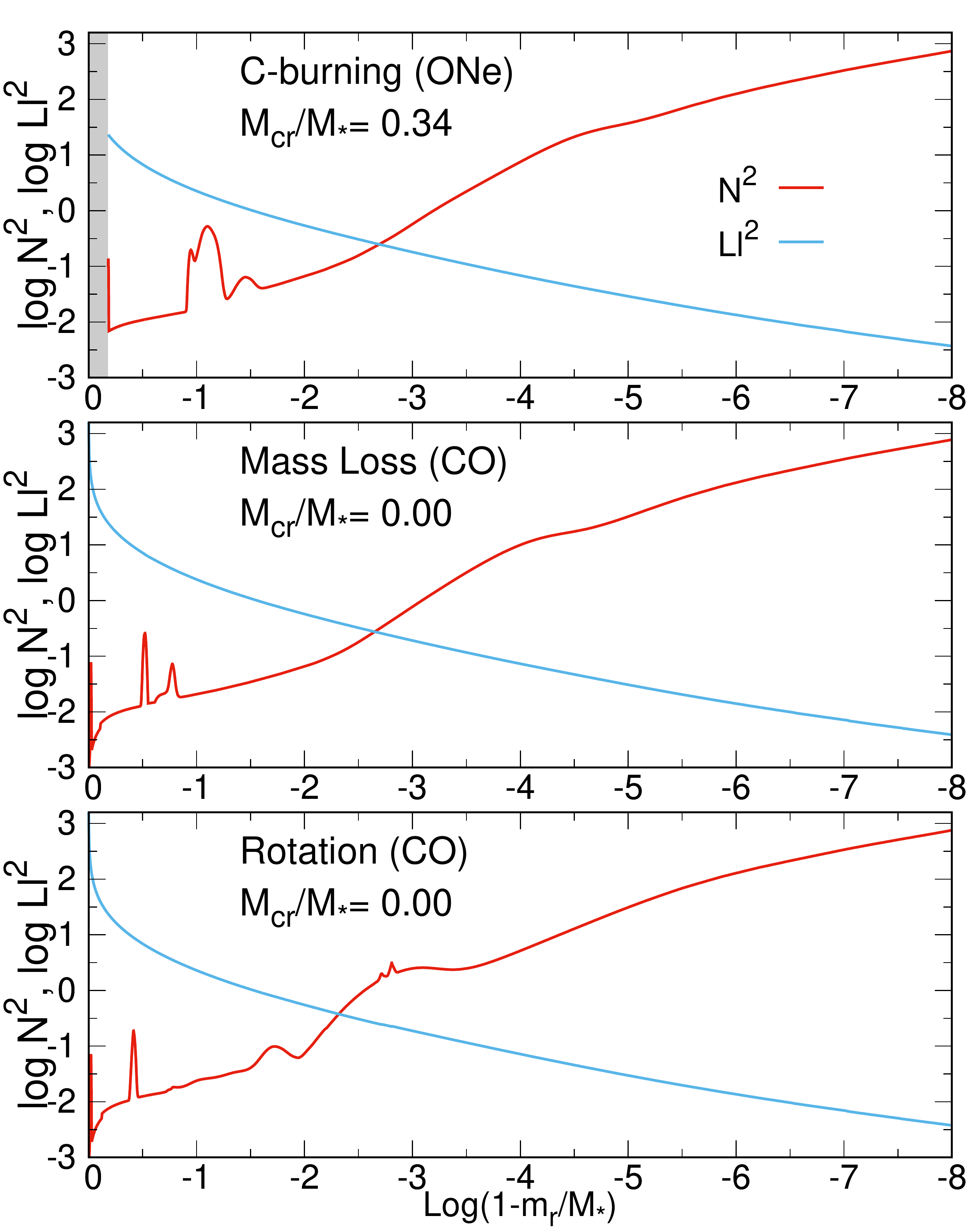}
        \caption{Same as Fig. \ref{bvf_30000.eps}, but for the models analyzed in Fig. \ref{perfiles_20000.eps}, with $M_{\star}= 1.156 M_{\odot}$ and $T_{\rm eff}\sim  20\,000$ K. The gray area in the upper panel corresponds to the crystallized part of the model.} 
        \label{bvf_20000.eps}
\end{figure}

\section{Pulsation results}
\label{pulsation_results}

\begin{figure}
        \centering
\includegraphics[width=1.0\columnwidth]{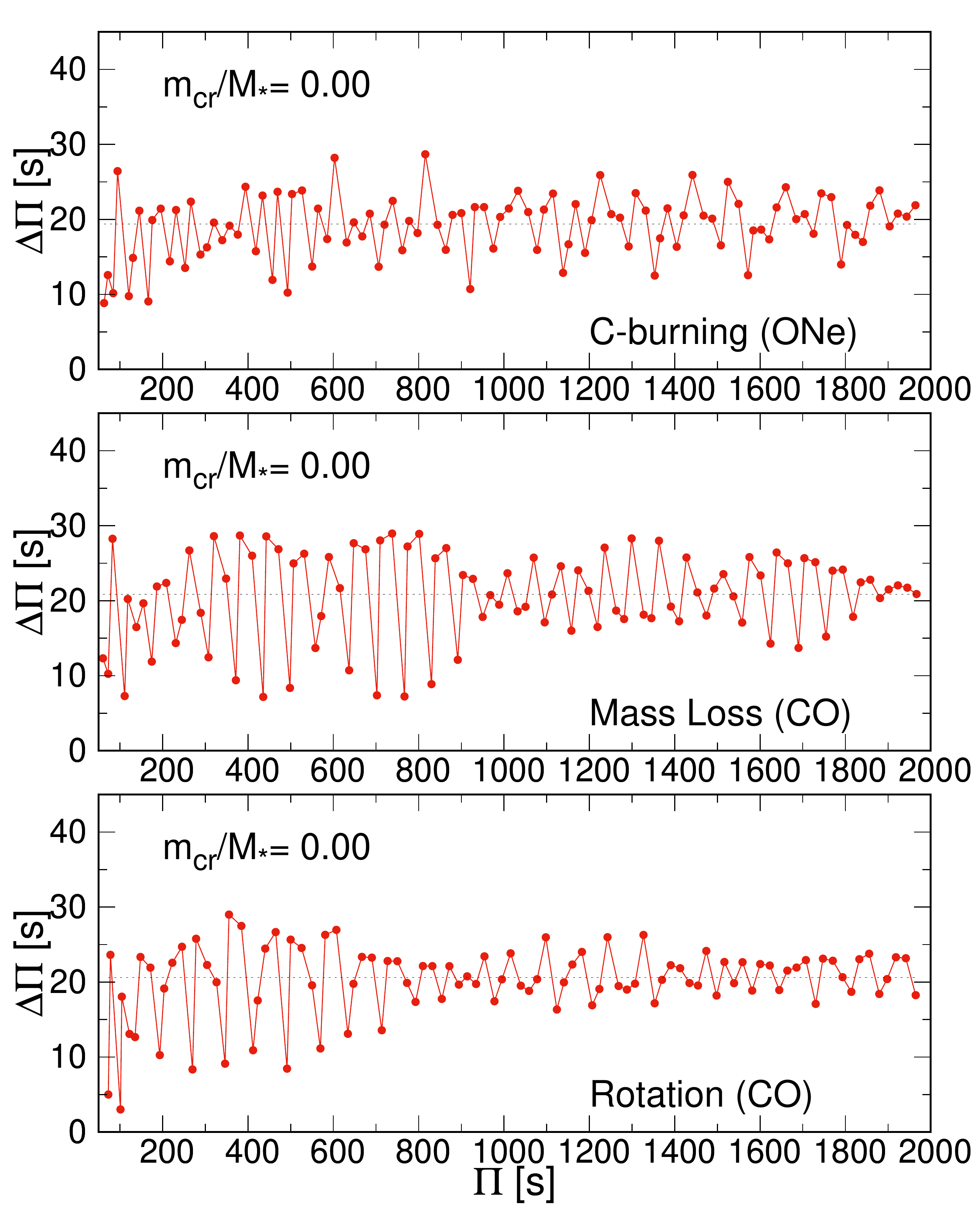}
        \caption{The forward period spacing ($\Delta \Pi$) in terms of the periods
  of $\ell= 1$ pulsation $g$ modes, corresponding to the models analyzed
 in Figs. \ref{perfiles_30000.eps} and \ref{bvf_30000.eps}, characterized by 
 $M_{\star}= 1.156 M_{\odot}$  and $T_{\rm eff}\sim 30\,000$ K. The horizontal 
 black-dotted line is the asymptotic period spacing.} 
        \label{dp_DB_30000.eps}
\end{figure}

\begin{figure}
        \centering
\includegraphics[width=1.0\columnwidth]{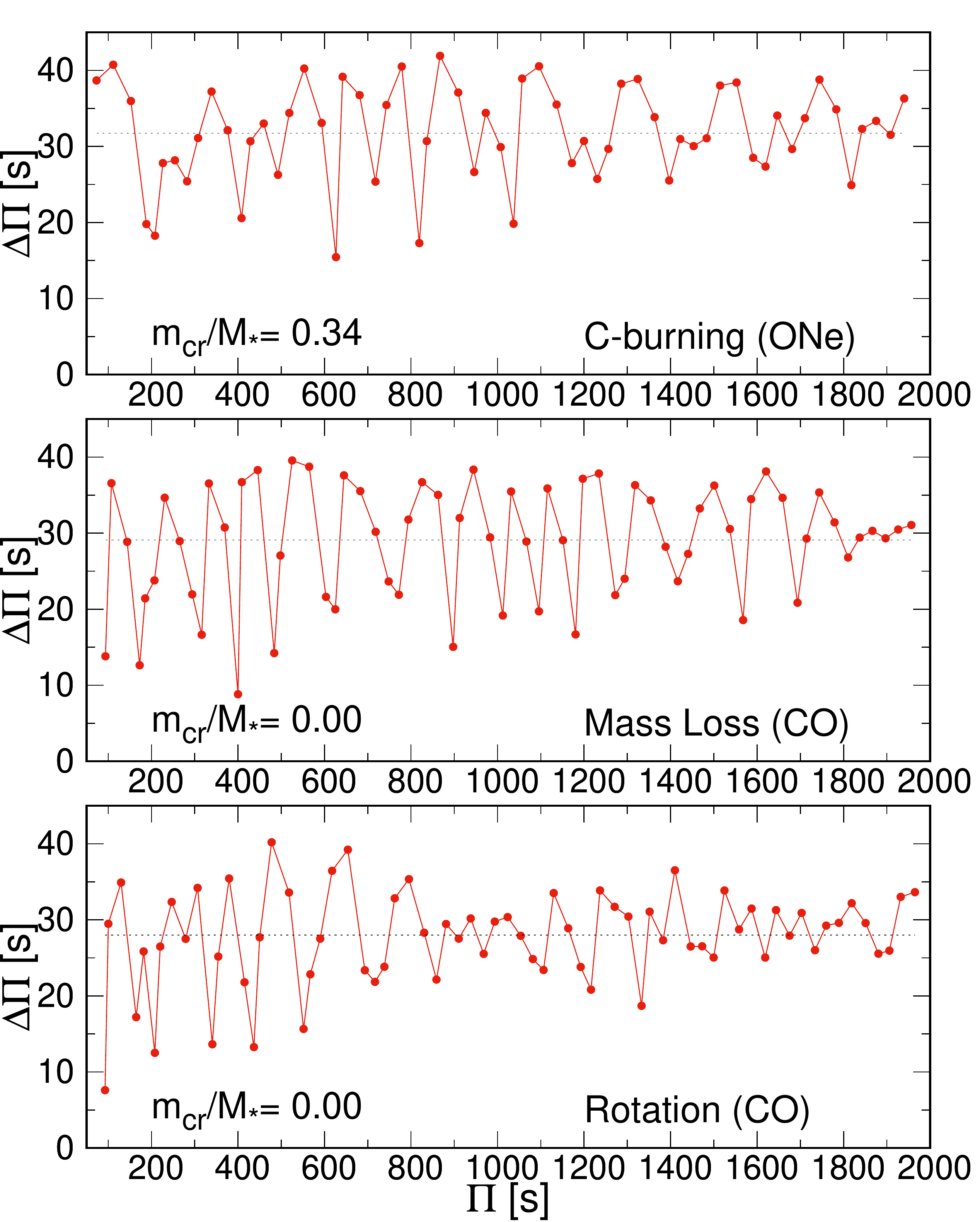}
        \caption{Same as Fig. \ref{dp_DB_30000.eps}, but for the models analyzed
 in Figs. \ref{perfiles_20000.eps} and \ref{bvf_20000.eps}, with 
 $M_{\star}= 1.156 M_{\odot}$  and $T_{\rm eff}\sim 20\,000$ K.} 
        \label{dp_DB_20000.eps}
\end{figure}

In this section, we compare the  pulsational properties of our  $1.156 
M_{\odot}$ and $1.29 M_{\odot}$ UMCO DB
WD  models that are the result of two single  scenarios  based  on rotation and reduced
mass-loss rates,  with
those predicted for the UMONe DB WD models resulting from off-center
C-burning  during   the  single  evolution  of   progenitor  stars
\citep{2010A&A...512A..10S}. For simplicity, we restrict ourselves to 
showing the results for two extreme temperatures: one corresponding to the 
hot edge, and the other typical of the cool edge of the DBV instability strip.

In   Figs.   \ref{bvf_30000.eps}   and
\ref{bvf_20000.eps}   we  compare   the  logarithm   of  the   squared
Brunt-V\"ais\"al\"a and  Lamb frequencies 
\citep[see][for their definition]{1989nos..book.....U} in  terms 
of the  outer mass
fraction of the  $1.156\,M_{\odot}$ DB WDs models at the blue
and red  edge of the DBV instability strip, respectively.  The shape  of the
Brunt-V\"ais\"al\"a  frequency  has  a  strong impact  on  the  
$g$-mode period
spectrum and mode-trapping properties of  pulsating WDs. The
Brunt-V\"ais\"al\"a frequency  of our ultra-massive  models 
does not exhibit  relevant features in  the outer layers.  This  is because
chemical diffusion   has  strongly
smoothed  out  the  $^{16}$O$^{12}$C$^{4}$He  interface (see  Figs.
\ref{perfiles_30000.eps}  and   \ref{perfiles_20000.eps}),  translating
into    a   very    smooth    shape    of   the    Brunt-V\"ais\"al\"a
frequency. However,  there exist dominant  features in the
run of  the Brunt-V\"ais\"al\"a frequency associated  to the innermost
chemical  transition  regions.  We note  that
different  bumps are  present  at different  locations, reflecting  the
location of the core chemical transitions  in each model.  

In Figs.  \ref{dp_DB_30000.eps} and  \ref{dp_DB_20000.eps} we show the
$\Delta \Pi  - \Pi$ diagram,  that is,  the separation of  periods having
consecutive  radial order  $k$ (the "forward  period spacing"  $\Delta \Pi
\equiv  \Pi_{k+1}-\Pi_k$) versus  the periods of  $\ell= 1$
pulsation $g$ modes, for our $1.156\,M_{\odot}$  DB WDs models at the blue
($T_{\rm eff} \sim 30\,000$ K) and  red ($T_{\rm eff} \sim 20\,000$ K)  
edge  of the DBV instability  strip,  respectively. Such  diagrams
constitute  a sensitive tool  for  studying  the  mode-trapping  properties  in
pulsating WDs.  Mode-trapping features, that are inflicted by bumps in
the  Brunt-V\"ais\"al\"a frequency,  manifest themselves in that  at a
given $T_{\rm eff}$ value,  the separation between consecutive periods
departs from the mean (constant)  period spacing. For the
stellar mass considered and for the complete range of effective temperatures 
of the instability strip, the
models  exhibit notable  mode-trapping  features  for the  whole
period  range. $\Delta  \Pi$ is  characterized  by maxima  and
minima, typical of WD models harboring one or more chemical interfaces.
These maxima  and minima represent  departures from a  constant period
separation,  which is  represented in  the figures  by the  asymptotic
period spacing  (horizontal black-dotted  line). 

The $1.156 M_{\odot}$ UMCO DB WD models we 
analyse here do not develop crystallization  during  the  DBV   instability  
strip,  allowing  the pulsation modes to "feel" the presence  of the chemical 
transition regions in the core, and  eventually producing the mode-trapping  
features in all of the models. This is in contrast with the situation encountered in the ultra-massive DA  WDs, which, at the evolutionary stages they pulsate, 
have their cores mostly crystallized. In the presence of crystallization, any  chemical interface located within the crystallized  region in each model has no
relevance  to the  pulsation properties  of  the $g$  modes,  since their 
eigenfunctions cannot penetrate  the solid
region of the stellar core. In the case of the $1.156 M_{\odot}$ UMCO DB WDs models 
studied here, the pulsations are able to probe 
from the surface to the very center of the star all along the instability strip.  
A notable feature of Fig. \ref{dp_DB_30000.eps} is that the pattern of forward 
period spacing for the models with CO core is quite different from 
that of the ONe-core WD model, particularly for modes with periods 
longer than $\sim 200$~s. So, in principle, for pulsating ultra-massive 
DB WDs with stellar masses close to $1.16 M_\sun$ evolving at the hot 
edge of the DBV instability strip, it could be possible to distinguish the different core chemical structures and compositions through pulsations, 
and thus provide clues about the evolutionary channels that led to their formation.  

The global characteristics of forward period spacing 
become similar for the $1.156 M_{\odot}$ models associated with the three scenarios when the 
they are close to the cool boundary of the DBV instability strip (Fig. \ref{dp_DB_20000.eps}). The only difference ---which is barely visible--- is related to the asymptotic period spacing, which slightly exceeds $\sim 30$ s in the case of the ONe-core WD model due to the crystallized region, compared to the cases of the UMCO WD models, that are not crystallized at all and have asymptotic period spacings slightly less than $\sim 30$ s. In practice, the mean period spacing is very difficult to measure because many periods with consecutive radial orders are needed. And it is even more difficult to  observationally estimate the tiny difference between the values of the mean period spacing of one formation scenario or another, as predicted by our theoretical 
computations. We conclude that, in the case of pulsating ultra-massive DB WDs with masses close to $1.16 M_\sun$ and temperatures near $T_{\rm eff} \sim 20\,000$ K, we would not be able to distinguish between the different types of structures and chemical compositions of the core, thus depriving the possibility of inferring their evolutionary origin.

\begin{figure}
        \centering
\includegraphics[width=1.0\columnwidth]{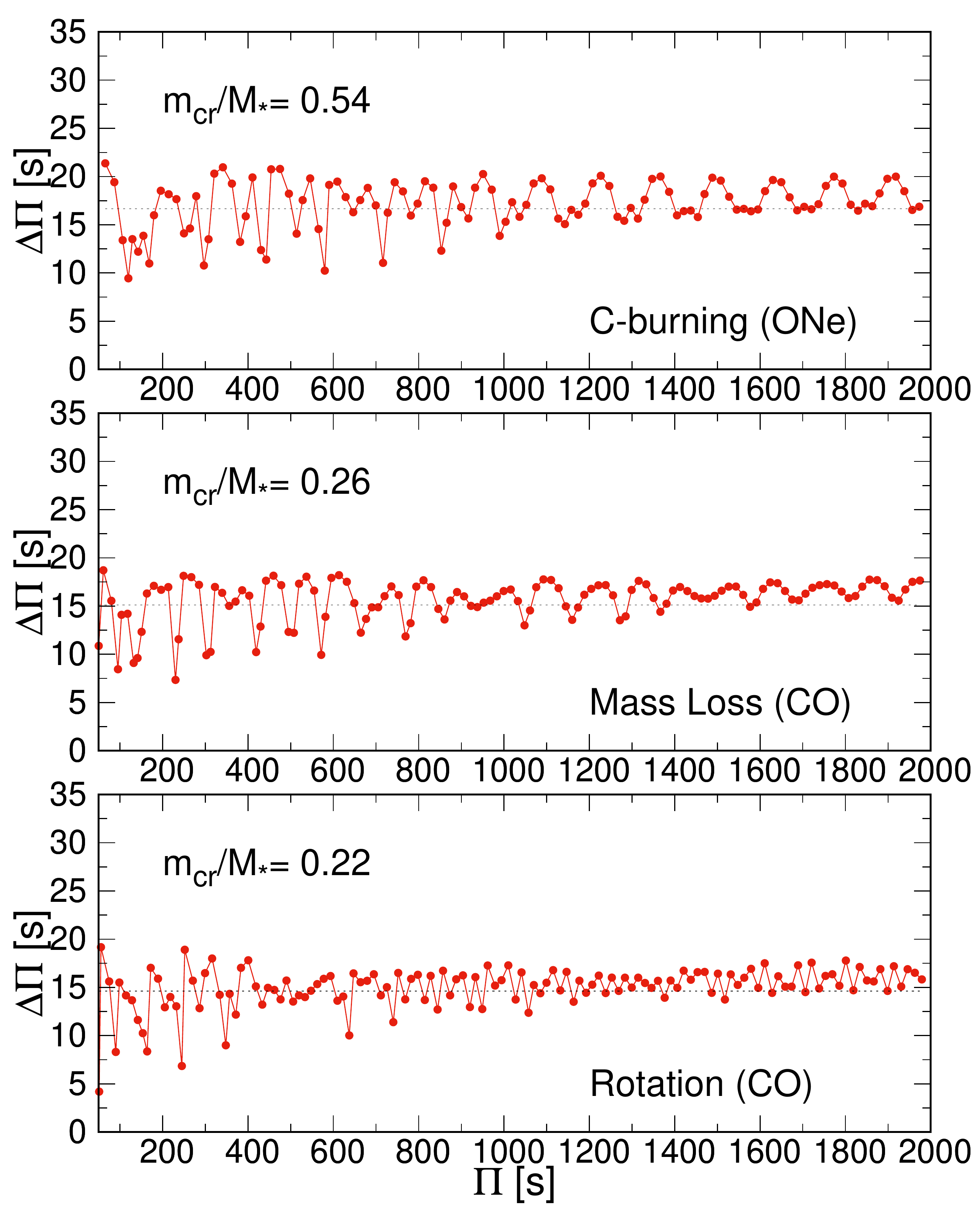}
        \caption{Same as Fig. \ref{dp_DB_30000.eps}, but for a stellar mass of $M_{\star}= 1.29 M_{\odot}$ ($T_{\rm eff}\sim 30\,000$ K).} 
        \label{dp_DB_30000-129.eps}
\end{figure}

\begin{figure}
        \centering
\includegraphics[width=1.0\columnwidth]{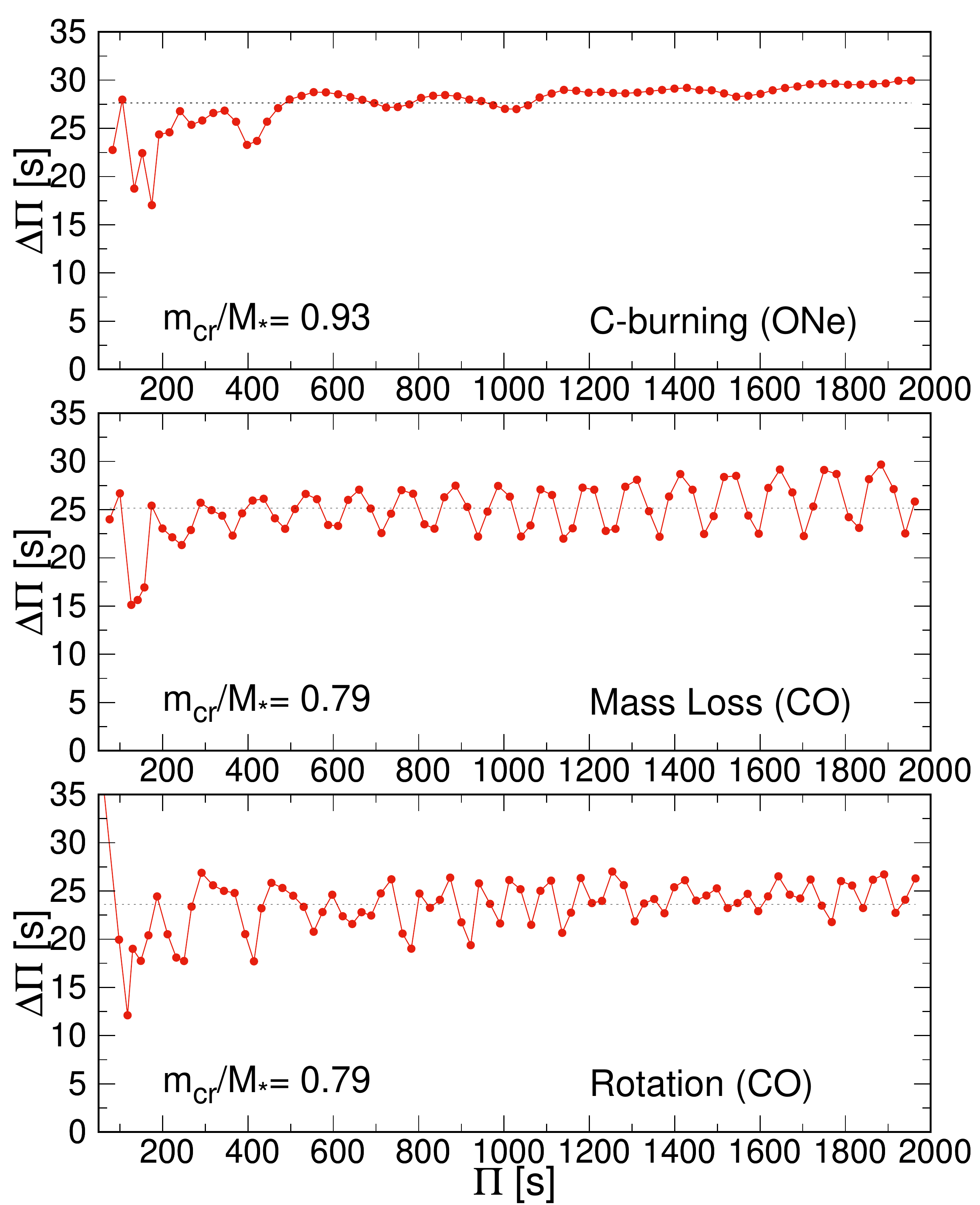}
        \caption{Same as Fig. \ref{dp_DB_30000-129.eps}, 
        but for $T_{\rm eff}\sim 20\,000$ K ($M_{\star}= 1.29 M_{\odot}$).} 
        \label{dp_DB_20000-129.eps}
\end{figure}

We turn now to the case of a more massive model sequence, that of $1.29 M_{\odot}$. Because of larger central densities, crystallization develops in the core of the models for 
all the formation scenarios during the DBV instability strip, even at its hot 
boundary. In Figs. \ref{dp_DB_30000-129.eps} and \ref{dp_DB_20000-129.eps} we depict the 
forward period spacing in terms of periods for models of $M_{\star}= 1.29 M_{\odot}$ at $T_{\rm eff}\sim 30\,000$ K and $T_{\rm eff}\sim 20\,000$ K, respectively. As can be seen, again at the high effective temperatures 
typical of the blue edge of the instability strip (Fig. \ref{dp_DB_30000-129.eps}),
the WD models that come from the three evolutionary scenarios all exhibit  
deviations from a constant period spacing due to mode trapping. 
We note that, although the mode-trapping amplitudes ---the magnitude
of the deviations from the constant period spacing---  are 
similar in the three models, it is nevertheless possible to 
distinguish different trapping cycles ---the intervals of periods between two consecutive minima of  $\Delta \Pi$--- in the case of the CO-core model 
with rotation, in relation to the other two scenarios. This difference could 
be, in principle, exploited from an observational point of view to distinguish the 
core chemical composition and the evolutionary channel of the star, 
provided that a large number of $g$ modes with consecutive $k$ values were detected 
in a $\sim 1.30 M_{\odot}$ DB WD star at high effective temperatures.

\begin{figure}
        \centering
        \includegraphics[width=1.0\columnwidth]{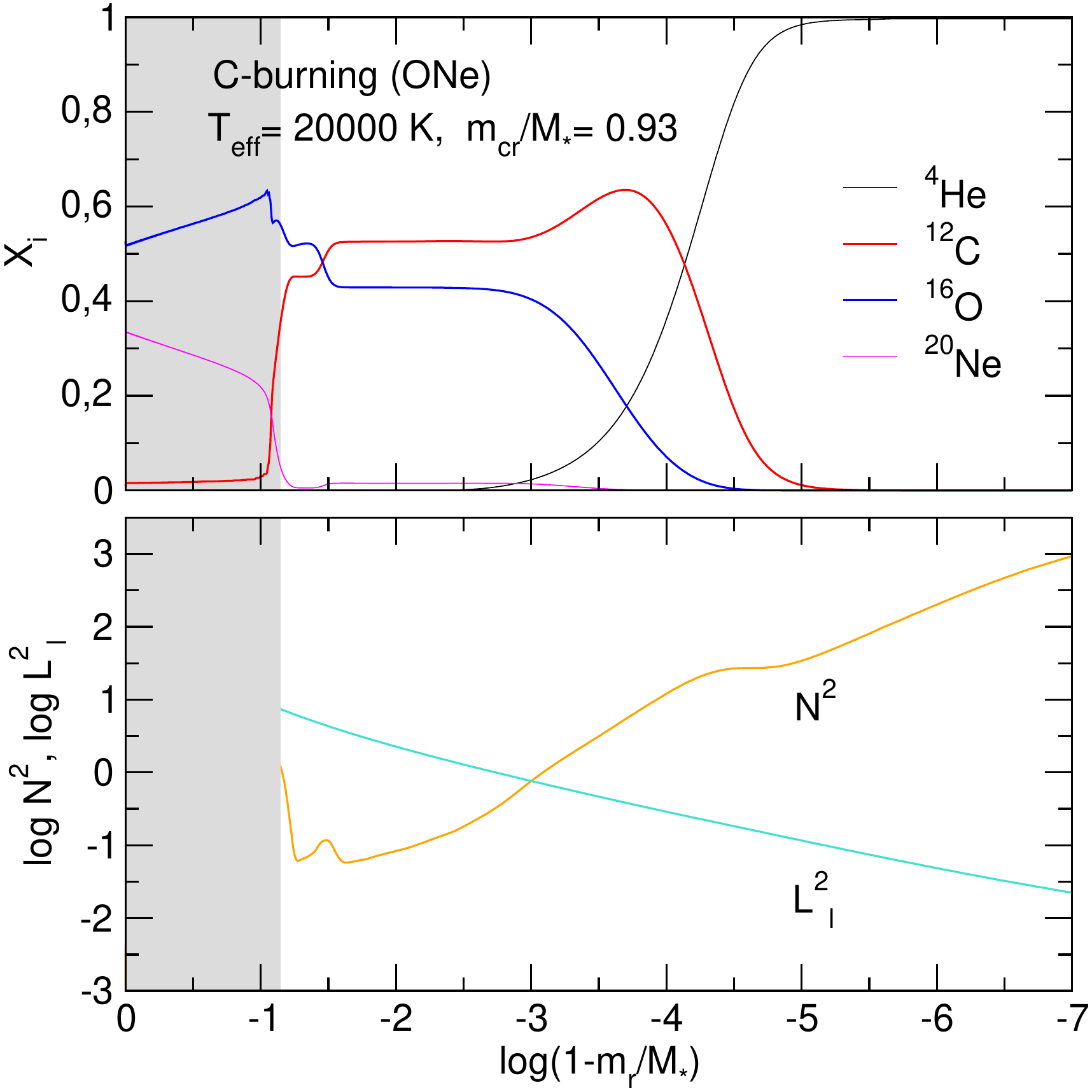}
        \caption{Upper panel: abundance by mass of $^4$He, $^{12}$C, $^{16}$O, and $^{20}$Ne versus the outer mass coordinate for the ONe-core $1.29M_{\odot}$ DB WD 
        model at $T_{\rm eff}\sim 20\,000$ K. Lower panel: logarithm of the squared Brunt-V\"ais\"al\"a and Lamb frequencies ($\ell= 1$).} 
        \label{x-bv-129-one-20000}
\end{figure}

At the other end of the DBV instability strip, near the red boundary 
(Fig. \ref{dp_DB_20000-129.eps}), the ONe-core WD model exhibits 
strongly weakened mode-trapping signals in the period spacing. 
This behavior can be 
understood as follows.  For the ONe-core WD model, that start to crystallize at  
$T_{\rm eff}\sim 39\,700$ K, the chemical interface located 
at $\log(1-m_r/M_{\star})\sim -1$ ---which is responsible for the mode 
trapping exhibited by this WD model--- 
ends up being contained in the crystallized part of the core by 
the time the model reaches 
the cool edge of the DBV instability strip. This can be seen in the upper panel of Fig. \ref{x-bv-129-one-20000}, corresponding to $T_{\rm eff} \sim 20\,000$ K.
This results in a very smooth Brunt-V\"ais\"al\"a frequency 
(see lower panel of Fig. \ref{x-bv-129-one-20000}), resulting in an almost constant period spacing (upper panel of Fig. \ref{dp_DB_20000-129.eps}). 
At variance with this, in the case of the CO-core models, the period 
spacing shows notorious mode-trapping features, which are due to 
the presence of spikes in the 
Brunt-V\"ais\"al\"a frequency that are still within the propagation cavity of the $g$ modes (that is, outside the crystallized regions). We conclude that, 
in the case of very massive ($\sim 1.30 M_\sun$) ONe-core pulsating DB WDs  near the red edge of the DBVs instability strip, the period spacing should be almost devoid of departures from a constant period separation, somewhat which 
would help to distinguish them from their CO-core counterparts.

\section{Summary and conclusions}
\label{conclusions}

In this paper, we have extended the scope of the analysis 
of \cite{2020arXiv201110439A}  by exploring the adiabatic pulsational  
properties of ultra-massive DB WDs resulting from  single-star 
evolution. Ultra-massive H-deficient WDs
are less frequent than H-rich objects, but at least a handful of them 
have been detected in the SDSS \citep{2013ApJS..204....5K, 2014A&A...572A.117R,
2020A&A...635A.103K,2020ApJ...901...93B}. In addition, an 
ultra-massive DB WD with an effective temperature well within the DBV instability strip  has been detected 
by \cite{2019ApJ...880...75R}  in   a  young  open
cluster, and a hot rapidly rotating DBA WD with a stellar mass of $1.33 M_{\odot}$ has been discovered by \cite{2020MNRAS.499L..21P}.
By means of the analysis of the period-spacing and mode-trapping features, 
we have shown that the pulsational properties of  the ultra-massive DB 
WDs are much more strongly dependent  on their formation scenario than   in the case of DA 
WD ones studied in \cite{2020arXiv201110439A}. This is 
due to the fact that DBV stars are much hotter than DAV stars, and therefore, 
their cores are substantially less crystallized.  As a result, $g$-mode pulsations 
in ultra-massive DBVs can penetrate much deeper in the star, thus 
carrying valuable information about the 
core chemical structure and composition. 

In particular, we expect that pulsating UMCO 
DB WDs 
%with stellar masses near $\sim 1.16 M_{\sun}$ 
at the hot boundary 
of the DBV instability strip ($T_{\rm eff} \sim 30\,000$ K) display 
mode-trapping features with larger amplitudes than those of their ONe-core 
counterparts (Fig. \ref{dp_DB_30000.eps}). In the case of very massive 
pulsating DB WDs ($M_{\star} \gtrsim 1.30 M_{\odot}$) we find that, 
if the stars are located near  the blue edge of the DBV instability domain, 
then it would be possible to differentiate the CO-core of a WD coming 
from the rotation scenario from the CO-core of a WD coming from the mass-loss 
scenario and the ONe-core of a WD resulting from the C-burning scenario
(Fig. \ref{dp_DB_30000-129.eps}). Admittedly, in order to make such a 
distinction, it would be necessary 
for the star to show many periods with consecutive radial orders.
On the other hand, if the very massive DB WDs are detected near the 
cool edge of the DBV instability strip ($T_{\rm eff} \sim 20\,000$ K, see Fig. \ref{dp_DB_20000-129.eps}), then we could have two different situations. 
If the stars exhibit substantial mode-trapping features, this might be 
reflecting that they are ultra-massive 
CO-core WDs. On the contrary, the absence of mode-trapping signatures would
be indicative that the stars have ONe cores.

We conclude that the eventual  detection of
pulsating ultra-massive DB WDs will constitute a much clear and unique
opportunity to discern the core composition and origin of the
ultra-massive WD population in general.  While pulsations have not been 
detected in any ultra-massive DB WD so far, there is now
an excellent prospect of observing pulsations in WDs in general, and 
in ultra-massive DB WDs in particular, thanks to ongoing space missions 
such as {\it TESS}, or space missions that will be operational in the next few years, 
such as {\it Cheops} \citep{2018A&A...620A.203M} and 
{\it Plato} \citep{2018A&A...620A.203M}.

\begin{acknowledgements}
We  acknowledge  the valuable suggestions  and comments of our
    referee, P.-E. Tremblay, that improved the original version of this paper. Part of  this work was  supported by  PIP 112-200801-00940 grant from CONICET,  by MINECO grants AYA2014-59084-P, and AYA2017-86274-P, by grant G149 from University of La Plata, and by  the AGAUR grant SGR-661/201. 
This  research has  made use of  NASA Astrophysics Data System.
\end{acknowledgements}

\bibliographystyle{aa}
\bibliography{ultramassiveCO}

%\newpage

%\begin{appendix}

\end{document}